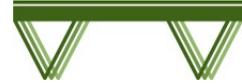

# IN-SERVICE PERFORMANCE AND BEHAVIOR CHARACTERIZATION OF THE HYBRID COMPOSITE BRIDGE SYSTEM – A CASE STUDY

John M. Civitillo
University of Virginia, USA

Devin K. Harris
University of Virginia, USA

Amir Gheitasi
University of Virginia, USA

Mark Saliba
University of Virginia, USA

Bernard L. Kassner
Virginia Center for Transportation Innovation and Research, USA

## ABSTRACT

The Hybrid Composite Beam (HCB) system is an innovative structural technology that has been recently used in bridge construction within the U.S. transportation network. In this system, the superstructure consists of a conventional reinforced concrete deck supported by Hybrid Composite Beams. Each beam is comprised of a glass-fiber reinforced polymer (FRP) box shell containing a tied parabolic concrete arch. Inclined stirrups provide shear integrity and enforce composite action between the HCBs and the concrete deck. This paper focuses on evaluating the in-service performance of a newly constructed HCB bridge superstructure located on Route 205 in Colonial Beach, Virginia. A live load test was conducted using tandem axle dump trucks under both quasi-static and dynamic conditions. Results obtained from the experimental investigation were used to determine three key behavior characteristics. Dynamic amplification and lateral load distribution were found to be reasonable in comparison to the assumed design values. The testing program also included internal and external measurement systems to help characterize the load sharing behavior of the HCB on an element level. The main load carrying elements are the deck in compression and the steel ties in tension, and the FRP shell did not act compositely with the internal components.

## 1. INTRODUCTION

### 1.1 Background

The Hybrid Composite Beam (HCB) system consists of a glass fiber reinforced polymer (FRP) box shell that encases a passively tied concrete arch. The tie reinforcement is unstressed prestressing strand that is integrated into the shell fabrication, while the arch is typically made of self-consolidating concrete that is able to fill the 100 mm thick arch profile created with foam inserts (Hillman 2008). Within the design, the concrete arch resists the internal compression forces due to self weight and additional construction loads, while the steel is intended to tie the arch together and carry internal tensile forces. Reinforcing bars oriented at a 45° and anchored within the arch section are distributed along the beam to provide composite action between the beam and a conventionally reinforced concrete deck. In this configuration, when the bridge is under service loads, the major load resisting components are the deck in compression and the steel in tension (Van Nosdall et al. 2013). The efficient use of materials, lightweight design, corrosion resistance, and application in accelerated construction applications are some of the advantages that make the HCB system an attractive alternative to conventional bridge designs (Hillman 2008).

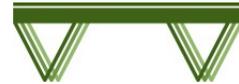

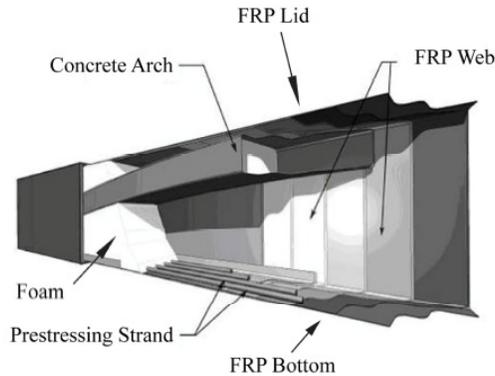

Figure 1: Hybrid Composite Beam Design (Hillman 2008)

## 1.2  Problem Statement

The Virginia Department of Transportation (VDOT) was interested in deploying this design technology in a bridge, however, the agency wanted verification of the structural behavior of the HCB. Thus, the purpose of this investigation was to quantify some of the unknown behavior of this relatively new bridge technology, and how this skewed configuration performs relative to existing bridge design principles and existing HCB applications with no skew. Of particular interest was the flexural lateral load distribution behavior, the element load sharing behavior, and the dynamic load amplification of a skewed HCB structure.

## 2. EXPERIMENTAL PROGRAM

This study examined the initial in-service behavior of a recently constructed HCB structure located on Route 205 over the Tides Mill Stream in Colonial Beach, VA. The construction involved replacing the previous concrete girders while maintaining the existing substructure. The HCB girders were 53 cm deep by 61 cm wide, and covered a 13.5 m clear span with a 19 cm cast-in-place composite deck. The girders were spaced at 1.2 m center-to-center, and the bridge measured 9.9 m transversely. The bridge carried two lanes of traffic, each with 0.9 m shoulders. The bridge spanned straight across a creek with two abutments on a 45° skew. In addition, the beams were made integral with the abutments by encasing the end sections of the beam in the abutment concrete. The in-service behavior of the system was evaluated through live-load testing.

## 2.1  Instrumentation Plan

Although this particular investigation focused on the initial in-service performance of the HCB bridge, the instrumentation used on this structure was designed to serve two functions: 1) to yield measured responses during load testing and 2) provide long-term monitoring capabilities for the HCB bridge system. Two separate and complementary data acquisition systems were used to achieve this goal: the first system used a module recently developed by Campbell Scientific Inc. (CSI) that is capable of dynamically measuring internally placed vibrating wire gages (VWG). The second system was a rapidly deployable wireless field testing system from Bridge Diagnostics Inc. (BDI) that measured externally mounted strain gages. Both systems were used during the load testing, but only the CSI system was intended for the long-term monitoring phase.

Vibrating wire gages were installed internally on three of the HCBs at various stages in the fabrication phases (Figures 2 and 3). The locations selected for installation were determined in collaboration with VDOT and were intended to measure the HCB system behavior during load testing, but also over the long-term for inspection and evaluation purposes. Along the length of the beam, gages were installed at midspan and at both quarterspan locations to accommodate the influence of the skew. Additional gages were installed at the eighth points to measure shear response, but were not the focus of this particular study. At each longitudinal location, VWGs were installed during production of the composite girders at the location of the tension steel prior to infusion process of the FRP shell. Non-corrosive acrylonitrile butadiene styrene 3D printed frames were designed to hold two additional VWGs. These frames were installed in companion locations within the arch conduit prior to the placement of the arch concrete. In addition, VWGs were affixed to the deck reinforcement at the same companion locations to provide a more comprehensive measure of the overall system behavior including the neutral axis location and level of composite action between the deck and girders.

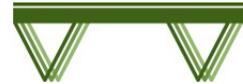

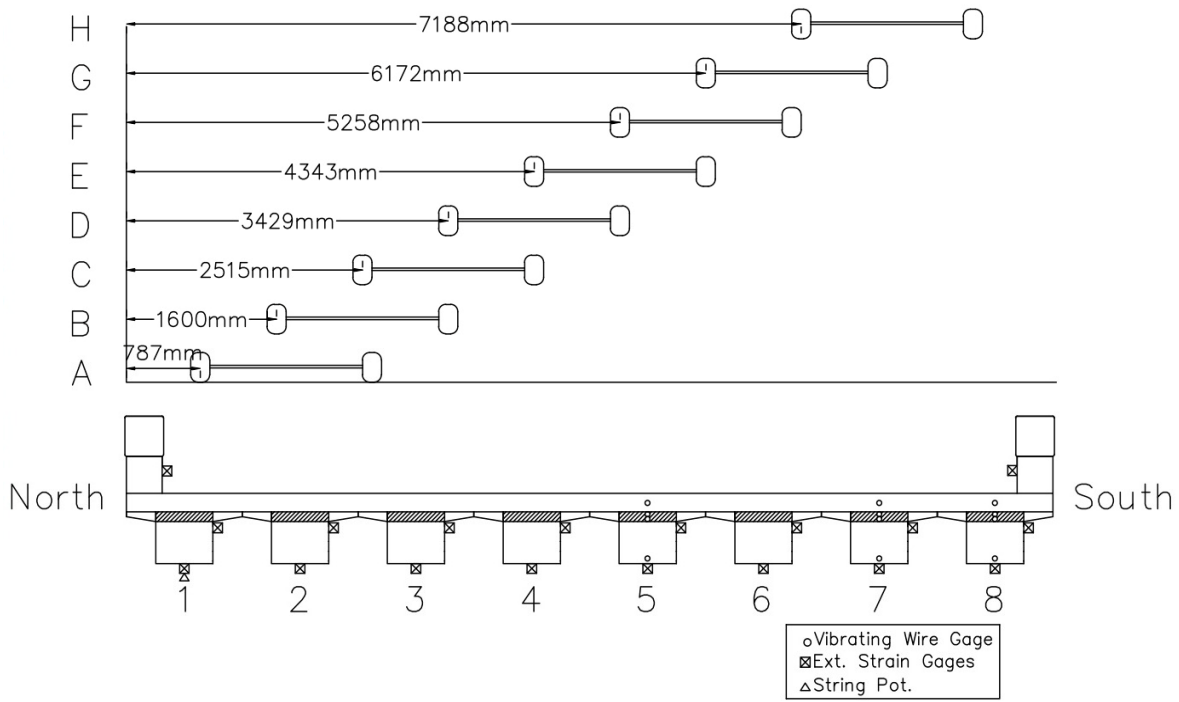

Figure 2: Instrumentation Layout and Test Truck Loading Configurations (Cross-Section View)

In conjunction with the internally placed gages, external strain gages were mounted on the FRP shell on the day of testing (Figures 2 and 3). These external gages were mounted on the bottom flange of each girder at midspan and the East quarterspan in order to capture distribution behavior of the bridge. Additional gages were mounted on the web of these girders at locations corresponding to the embedded VWGs in the arch (See Figure 2). These gages on the web gave an external strain profile that could be compared to the internal strain profile and help evaluate the load sharing behavior of the individual HCB components. Finally, two additional gages were mounted at midspan on the parapets to capture their contribution to the overall system behavior.

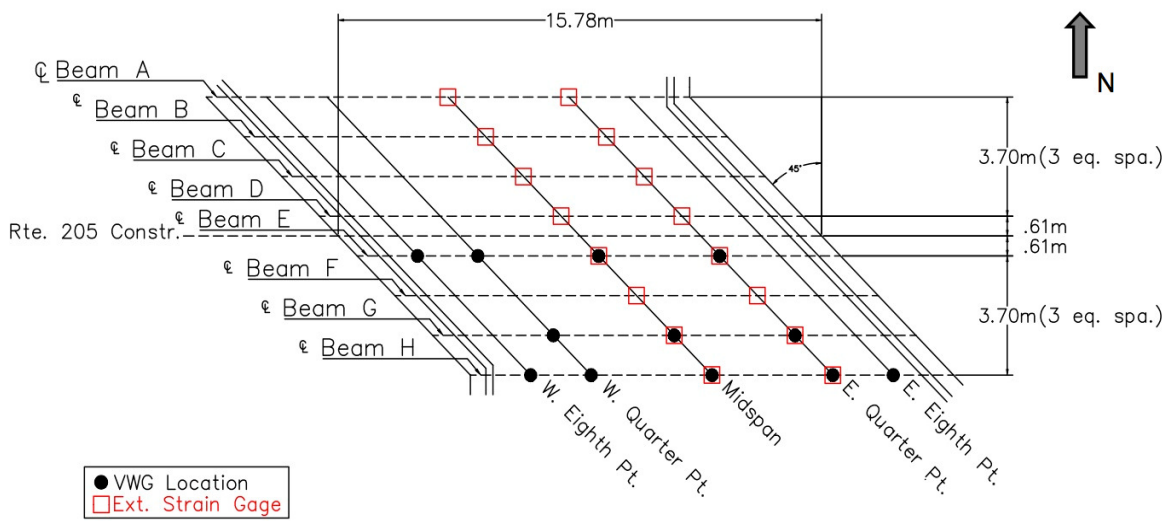

Figure 3: Instrumentation Bird's Eye Overview

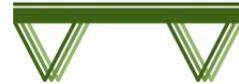

## 2.2  Load Test

The load testing was performed by driving a load vehicle east to west across the bridge at predetermined transverse positions (Figure 2). The load truck used was a VDOT tandem-axle dump truck loaded with gravel, weighing 232 kN (80 kN front axle and 76 kN per rear axle). The truck was positioned in eight different transverse locations in order to elicit a maximum response from each of the eight girders during quasi-static testing, which consisted of driving the test vehicle across the bridge at a slow speed (<8 kph). Each transverse position was conducted three times. During the final repetition, the truck remained parked at midspan for about 1 minute to allow for some of the non-dynamically measured VWGs to collect data under a true static loading. These static tests provided critical information regarding the system-level load sharing behavior as well as internal composite load sharing behavior of the HCB system.

In addition to the static testing, two dynamic tests were performed using a comparable truck with a slightly lighter load (225 kN gross). The dynamic tests were performed by driving the truck across the bridge near the posted speed limit (between 55 and 72 kph) 2 or 3 times each at transverse locations 'B' and 'D' on Figure 2. These two positions were selected due to their proximity to the three girders with internal VWGs. The static data provided critical information on the dynamic load allowance inherent to the HCB bridge system.

## 3. RESULTS

The load testing program yielded critical information for both the bridge system level behavior of an HCB bridge as well as the internal load sharing behavior at the element level of an individual beam. Results from the testing program allowed for the determination of flexural lateral load distribution behavior and dynamic load allowance at the system level as well as the relative contributions of the composite cross-section at the element level. These results are presented in this section.

### 3.1  Flexural Lateral Load Distribution

Flexural lateral load distribution describes the relative load sharing phenomena that occurs in a complex bridge system or the fraction of live load that is resisted by an individual member in a girder bridge system. The fractional representation is generally considered a simplification of complex two-way bridge system interactions down to a one-way behavior representation and is typically used for member design. In this study, the flexural lateral load distribution behavior was analyzed to help evaluate the in-service behavior of the HCB bridge because such in-situ performance is critical to the end user and there are very few HCB bridges currently in use.

Figure 4 illustrates the transverse longitudinal strain profiles at midspan for each of the load configurations. As would be expected in a typical load test, the girders directly under the load truck experience the greatest strain response, while the load diminishes over the neighbouring girders. It would be expected that an exterior girder might exhibit greater strain because it is unable to distribute load to a more exterior member, but in fact the strains in the exterior girder are always less than the most heavily loaded interior girder. It can also be seen that in each case where the load truck is near the barriers (cases 'A', 'B', 'G', and 'H'), the maximum strain response occurs not in the nearest exterior girder, but rather in the first interior girder. This behavior indicates a significant stiffening influence provided by the barriers to these outer girders. In a companion study by Ahsan (2012) that evaluated a full-scale, three girder HCB bridge of similar design, a similar distribution response was observed for the interior girder, but the exterior stiffening effects were not present due to the lack of parapets in their tests.

Equation 1, below, is commonly used to describe the flexural lateral load distribution response, where strains at midspan are the measured inputs. The average distribution factor for each of the load configurations are presented in Table 1. Observe that the controlling distribution factor for the exterior girder occurred for load case 'H', while the load case 'A' controlled for the interior girder. The distribution factors from the field test (especially the exterior girders) are lower than the results from the study by Ahsan (2012), indicating a more uniform load sharing phenomenon than what was found in the full-scale laboratory testing. It should be emphasized that from the AASHTO LRFD and the AASHTO Standard Specification do not contain provisions for this type of system, but when considering the anticipated HCB element behavior, it would be expected that the system behavior might mimic that of a conventional slab-girder bridge system such as a concrete deck on reinforced concrete girders (AASHTO Type

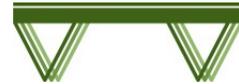

A) or a concrete deck on box girders (AASHTO Type B). When comparing the measured distribution behavior to these systems that include code-specified adjustments for skew, Type A and Type B designations yield conservative estimates for exterior girders, but the interior distribution values are non-conservative compared to traditional precast distributions.

[1]     $$DF_i = \frac{\varepsilon_{max\,i}}{\sum\limits_{i=1}^{\#\,girders} \varepsilon_{max\,i}} \cdot \left(\#\,of\,Trucks\right)$$

Table 1: Summary of Flexural Distribution Factors

|  | | Max. Exterior | Max. Interior |
|---|---|---|---|
| | Controlling | 0.225 | 0.349 |
| | Load Case 'A' | 0.188 | **0.349** |
| | Load Case 'B' | 0.135 | 0.316 |
| | Load Case 'C' | 0.094 | 0.256 |
| | Load Case 'D' | 0.059 | 0.220 |
| | Load Case 'E' | 0.051 | 0.199 |
| | Load Case 'F' | 0.103 | 0.230 |
| | Load Case 'G' | 0.164 | 0.281 |
| | Load Case 'H' | **0.225** | 0.298 |
| 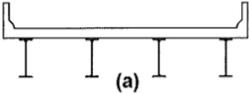 | AASHTO LRFD type 'A' CIP concrete deck steel beams | 0.247 | 0.360 |
| 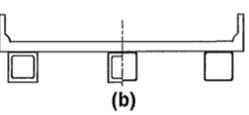 | AASHTO LRFD type 'B' CIP concrete deck closed steel or precast boxes | 0.288 | 0.306 |
| | AASHTO Standard Specification | 0.371 | 0.371 |
| | Virginia Tech (Ahsan 2012) | 0.390 | 0.360 |

Figure 4 also displays the internal VWG strain readings at the level of the strands for some girders for comparison to the external measurements. These internal strains show some agreement with the external FRP strain readings at lower strain values, but show less agreement at higher strain values. This response from the internal gages seems to contradict the measurements from the external gages because of the lack of response from the edge stiffness due to the parapet walls. Each of the tests from load cases 'E' through 'H' show that the internal strain readings are higher than the strains in the first interior beam, while the external FRP strains are lower in the fascia girder versus the first interior member. The VWG strains do not consistently match the trend or magnitude of strain. The cause of this non-correspondence is unknown but may be attributed to the non-composite behavior of the HCB components, as discussed below. This is reinforced by an observation of the strains near the exterior of the bridge near the parapet walls where there is a lower stiffening effect on the internal tension strands than the FRP shell. It is important to note that the main tension load carrying component is the internal tension steel, yet the distribution behavior has been derived from the external FRP strain readings, and may not represent the distribution behavior in the steel itself.

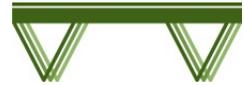

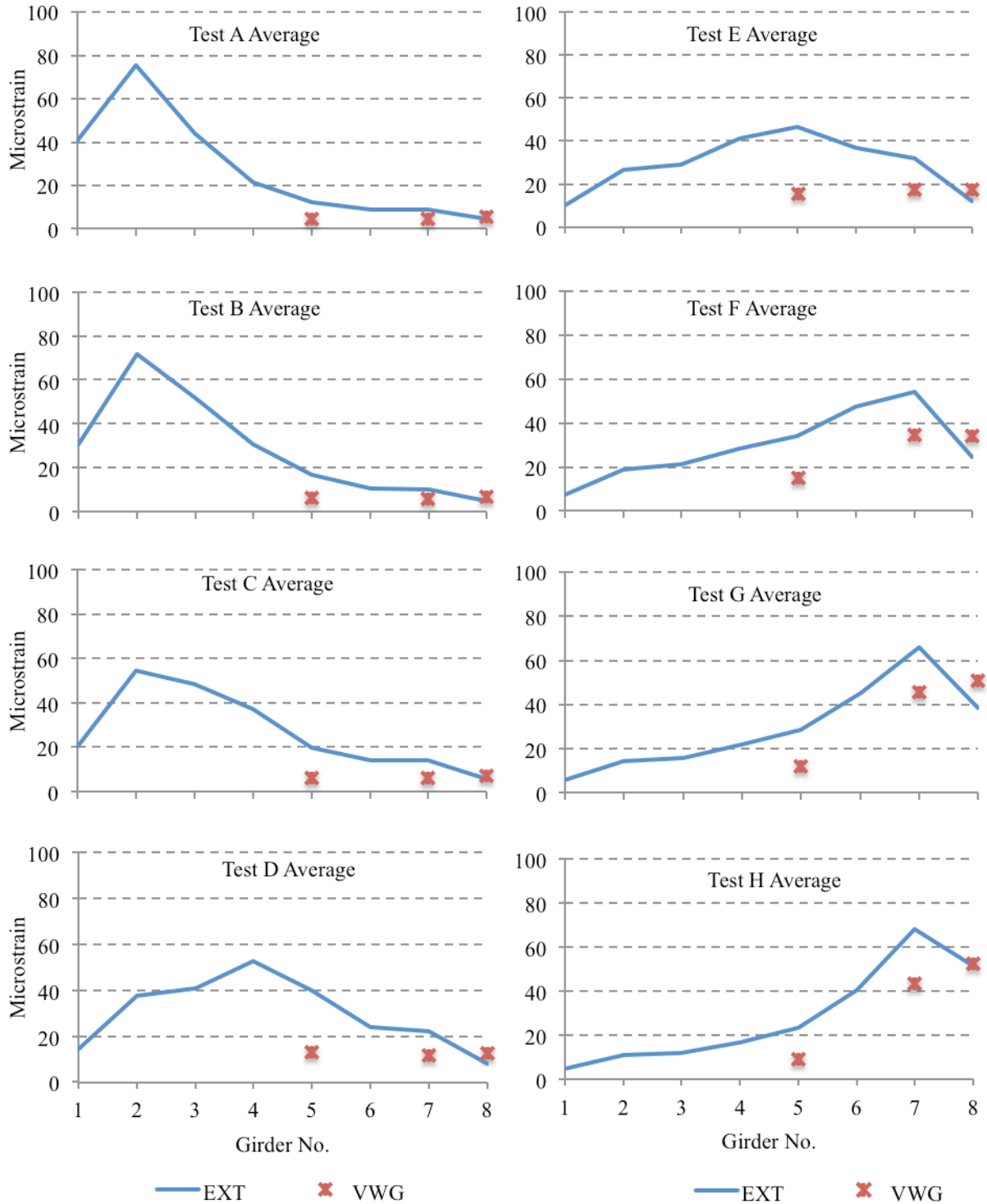

Figure 4: Transverse Distribution of Midspan Tension Flange Strain Versus Girder Number

## 3.2 Element Load Sharing Behavior

The HCB system is a composite system that is constructed in multiple stages and as a result, an understanding of the internal element level load sharing behavior is critical. An understanding of this internal load sharing behavior is also essential for maintenance and decision-making processes as the system ages. The internal and external instrumentation allowed for the measurement of the strain profile through the depth of girders 5, 7, and 8 during loading. The data provided critical information on the load sharing behavior between the concrete arch, FRP shell, and reinforcing steel, as well as the location of the neutral axis of the composite cross-section. One critical note is that while the HCB is designed as a tied arch system, once the deck is cast, the neutral axis occurs in the deck. Thus, the entire HCB is in tension at midspan under superimposed dead load (Van Nosdall et al. 2013). Figure 6 illustrates the strain profiles through the depths of girders 5,7, and 8 for load cases 'F', 'G', and 'H'. These load cases were selected because the load truck was in close proximity to the three girders that maintained both internal and external instrumentation, and thus yielded the most relevant results.

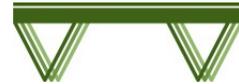

These thru-depth strain profiles show that the average strain profile of the VWGs yield locations for the neutral axis as the strains transition from tension in the HCB to compression in the concrete deck. While there appears to be some level of composite action with the deck, it is evident that the strain profile through the depth is not consistently linear as might be expected for full composite action. As an example, the tensile strains in the arch section of girder 5 exceed those in the tensile strand level. These high tensile strains in the bottom of the arch support Ahsan's (2012) local thin arch bending theory. It has been proposed that the arch acts non-compositely with the FRP webs and when a concentrated load is applied, the cross-section compresses, inducing flexure in the concrete arch.

A representative time-lapse plot of strains through the depth of a girder is shown in Figure 5 and shows that the two companion locations show some agreement. Both the arch level gages and tension level gages track in trend, but not magnitude. This is apparent in Figure 6, and again suggests the FRP shell is acting non-compositely with the internal load carrying components. It is also interesting to note the significant dip in strain in the arch gages before reaching their peak strain. The cause of this phenomenon is most likely due to the influence of two separate axle groupings that elicit two maximum responses, where the rear axle registers the maximum response.

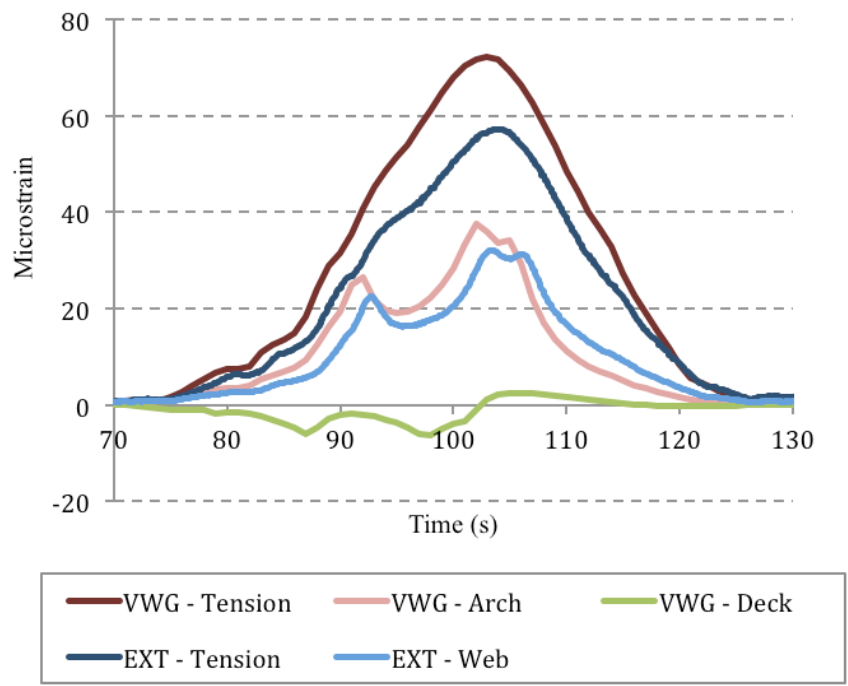

Figure 5: Representative Midspan Time-Lapse Strain Comparison
Girder 8 Response for Load Case H, Run 1

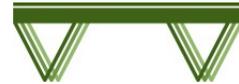

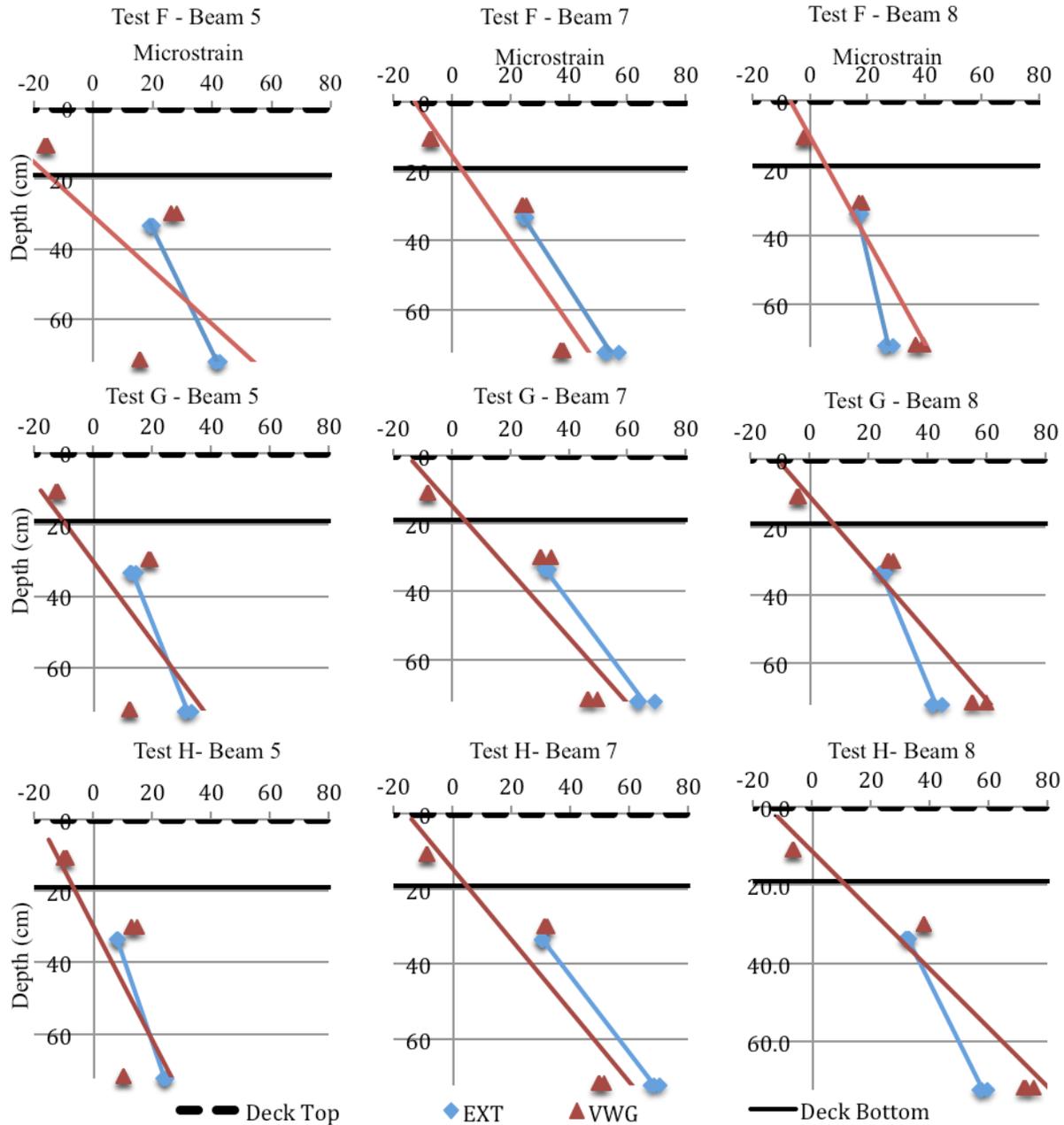

Figure 6: Strain Profiles through Midspan Depth of Truck Locations F Through H

## 3.3 Dynamic Load Allowance

The dynamic load allowance (DLA) or impact factor accounts for the amplification of the design static live load due to a moving vehicle crossing a bridge. This amplification is influenced by a number of factors, including bridge mass and flexibility, vehicle mass and suspension, roadway roughness, and vehicle speed. The dynamic amplification inherent to the HCB is critical because the system is relatively lightweight and more compliant than a traditional concrete or steel girder system. Load test results from a static load case and its corresponding dynamic load test were used to determine the dynamic load allowance for load cases 'B' and 'D' with the relationship presented in Equation 2. These cases were selected to align with results from the static load cases but also allow for safe travel of the load trucks across the bridge at speed. The original test plan included efforts to measure the VWGs dynamically, however, this effort proved unsuccessful and only external response was recorded. The amplification can be observed from the superposition of strains observed during the static and dynamic tests (Figure 7).

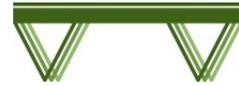

[2] $\mathrm{DLA} = \dfrac{\varepsilon_{dynamic} - \varepsilon_{static}}{\varepsilon_{static}} \cdot 100\%$

The maximum dynamic amplification response for each of the two load truck positions tested under dynamic conditions is presented in Table 2 along with the design values from the AASHTO LRFD and Standard Specification (AASHTO 2002 and 2012). These measured values represent the amplification observed in the most heavily loaded girders for the respective load cases. From the results it is evident that the average dynamic amplification is comparable to those prescribed for design, but there is the potential for larger amplification as demonstrated by the maximum responses observed in the most heavily loaded girders. In the maximum dynamic amplification cases below, the maximum response registered in girder 4 for load case 'B' was 46%, while a response of 33% was registered in girder 7 for load case 'D'.

Table 2: Dynamic Load Allowance

|  | Impact Factor |
|---|---|
| Load Case 'B' Maximum Response (Girder 4) | 46% |
| Load Case 'D' Maximum Response (Girder 7) | 33% |
| Measured Average (All Girders) | 23% |
| AASHTO LRFD 3.6.2.1 | 33% |
| AASHTO Standard Spec | 30% |

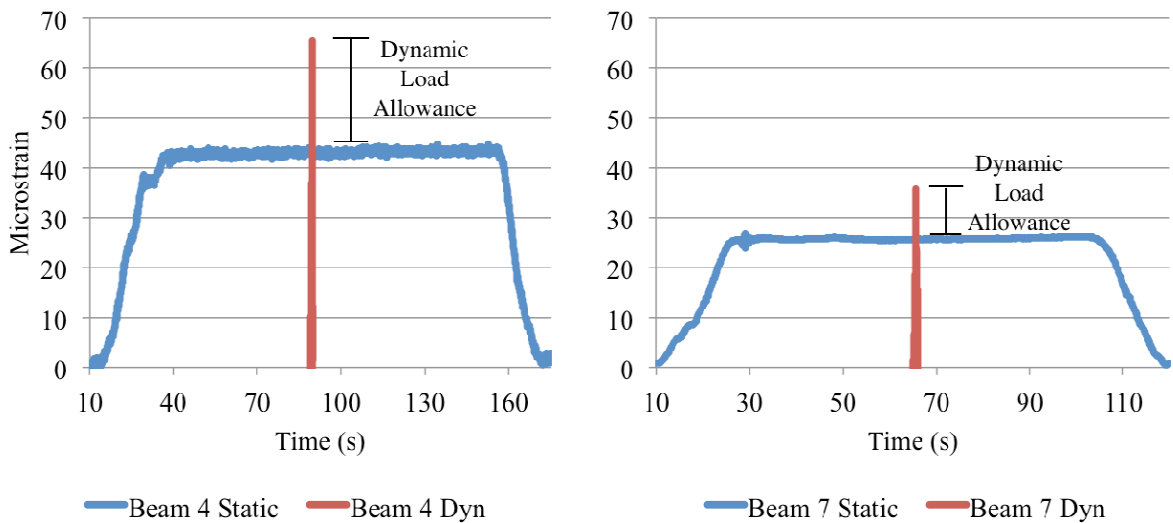

Figure 7: Maximum Dynamic Amplification Response Under Load Case 'B' (Left) and Load Case 'D' (Right)

## 4. CONCLUSIONS

This study was comprised of an in-service live load test of a HCB bridge system skewed at 45° over a 13.5 m clear span. This test was conducted to validate the results of a companion laboratory study performed by Ahsan (2012), and focused on three critical bridge behavior characteristics; lateral load distribution, dynamic load allowance, and internal load sharing behavior of all HCB components. This was accomplished through an extensive instrumentation program that embedded VWGs through the depth of the composite section at various lateral and transverse locations, as well as externally mounted strain gages at companion locations. The load test included both quasi-static and dynamic testing, and varying the transverse position of the load vehicle yielded sufficient data to characterize the bridge behavior.

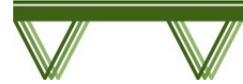

The findings of the study are summarized as follows:

- Distribution behavior determined from the externally mounted strain gages was consistent with expected trends. The highest strains were registered directly under the load vehicle, and dissipated elsewhere. The parapet walls offered a significant stiffening contribution to the fascia girders. The controlling experimental distribution factors confirmed Ahsan's (2012) laboratory work, and were comparable to traditional AASHTO type 'A' girder bridge values.
- It was found that the average dynamic load allowances registered were on par with the assumed design values, despite the observation of high levels of apparent compliance on the day of the load test. It was shown that there is potential for high dynamic amplifications, and more research may be warranted in this area.
- The internal strain profiles confirm the assumed neutral axis, but show a slight non-linear trend, proving that the FRP shell does not act compositely with the internal HCB components. The high tensile strains in the bottom of the arch confirm Ahsan's (2012) local arch bending theory.

Follow up activities to this study may include focused secondary load tests, long-term monitoring of the internal VWGs and an evaluation of potential NDE techniques for an inspection program.